\documentclass[preprintnumbers,article,amsmath,amssymb,floatfix,10pt,prd,onecolumn,superscriptaddress,nofootinbib]{revtex4}
\usepackage[colorlinks=true, pdfstartview=FitV, linkcolor=blue, citecolor=red, urlcolor=magenta]{hyperref}
\usepackage{bbm}
\usepackage{amsfonts}
\usepackage{mathrsfs}
\usepackage{latexsym}
\usepackage{epsfig}
\usepackage{epstopdf}
\usepackage{epstopdf}
\usepackage{graphicx}
\usepackage{amssymb}
\usepackage{amsmath}
\usepackage{dcolumn}
\usepackage{bm}
\usepackage{float}
\usepackage{color}
\usepackage{comment}
\usepackage{xcolor}
\begin{document}
\title{\bf Novel Gravastar Solutions: Investigating Stability, Energy, and Entropy in the Presence of Cloud of Strings and Quintessence}

\author{Faisal Javed}
\email{faisaljaved.math@gmail.com}\affiliation{Department of
Physics, Zhejiang Normal University, Jinhua 321004, People's
Republic of China}

\author{Ji Lin}
\email{linji@zjnu.edu.cn(Corresponding Author)}\affiliation{Department of Physics,
Zhejiang Normal University, Jinhua 321004, People's Republic of
China}

\begin{abstract}
Gravastars, theoretical alternatives to black holes, have captured
the interest of scientists in astrophysics due to their unique
properties. This paper aims to further investigate the exact
solution of  a novel gravastar model based on the Mazur-Mottola
(2004) method within the framework of general relativity,
specifically by incorporating the cloud of strings and quintessence.
By analyzing the gravitational field and energy density of
gravastars, valuable insights into the nature of compact objects in
the universe can be gained. Understanding the stability of
gravastars is also crucial for our comprehension of black holes and
alternative compact objects. For this purpose, we presents the
Einstein field equations with the modified matter source and
calculate the exact solutions for the inner and intermediate regions
of gravastars. The exterior region is considered as a black hole
surrounded by the cloud of strings and quintessence, and the
spacetimes are matched using the Darmoise-Israel formalism.
 {An investigation is conducted on the stability of
gravastars using linearized radial perturbation. Additionally, the
proper length, energy content, and entropy of the shell are
computed. The stability of gravastars is positively correlated with
the enhancement of the cloud of strings parameter, while it is
negatively correlated with the growth in the quintessence field
parameter.} The paper concludes with a summary of the findings and
their implications in the field of astrophysics and cosmology.\\ \\
\textbf{Keywords:} Gravastar; cloud of strings; quintessence field;
stability analysis.

\end{abstract}

\maketitle

\date{\today}
\section{Introduction}\label{sec:1}

Astrophysical objects known as black holes, which are the
mathematical solutions to the Einstein equations, are widely
acknowledged. Two significant observations about black holes in our
universe were recently reported. The first is the discovery of
gravitational waves coming from black hole binaries
\cite{1m},\cite{2m}, and the second is the discovery of photographic
proof of black holes at the centers of our galaxy's M87 and Sgr A*
\cite{3m},\cite{4m}. But distant observers should be able to observe
phenomena that happen outside the event horizon or even before they
form. As a result, it is still unclear whether an astrophysical
object has an event horizon. Several authors have suggested that the
gravitational collapse of a massive star could make the densest
celestial objects other than black holes. To address this idea,
Mazur and Mottola \cite{5m} introduced a new theory of collapsing
stellar objects known as gravitational vacuum stars, or
``gravastars", which incorporates the expanded concept of
Bose-Einstein condensation in the gravitational system. Gravastar
has been proposed as a black hole substitute that also takes into
account quantum effects. The gravastar model is believed to offer a
solution to issues associated with traditional black holes, while
also meeting all theoretical requirements for a stable endpoint of
stellar evolution. This theory suggests that quantum vacuum
fluctuations play a significant role in collapse dynamics, leading
to a phase transition that results in a repulsive de Sitter core
that balances the collapsing body and prevents the formation of a
horizon (and singularity) near the bound of $r=2m$. However, this
phenomenon occurs very close to the limit, making it challenging for
an outsider to differentiate between a gravastar and a true black
hole.

Gravastars exhibit a distinctive structure characterized by three
distinct zones. The internal region ($r\geq 0, r< r_{1}$) consists
of an isotropic de Sitter core governed by the equation of state
(EoS) $\rho=-p$. The external zone ($r> r_{2}$) corresponds to the
vacuum and is described by the Schwarzschild geometry, with an EoS
of $p=\rho=0$. Separating these internal and external regions is a
thin shell ($r_{1}<r< r_{2}$) of stiff matter having EoS $\rho=p$,
where $r_{1}$ and $r_{2}$ depict the internal and external radii of
gravastar. After the proposal of Mazur and Mottola, enormous
discussion has been done on gravastar. Visser and Wiltshire \cite{3}
analyzed the stable structure of gravastars and examined that
distinct EoSs yield the dynamically stable configuration of
gravastars. Carter \cite{4} presented novel exact gravastar
solutions and observed the impact of EoS on different zones of
gravastar geometry. Bili\'{c} et al. \cite{5} obtained the
gravastars solutions by considering the Born-Infeld phantom instead
of de Sitter spacetime and observed that at the center of stars,
their findings can manifest the dark compact configurations.

The role of energy limits within the gravastar shell and its
stability were investigated by Horvat and Ilijic \cite {6} using
radial perturbations and the sound speed on the shell. Various
researchers \cite{7}-\cite{12} have explored the internal structure
of the gravastar using different equations of state. Lobo and
Arellano \cite{13} built several gravastar models that incorporated
nonlinear electrodynamics and discussed specific characteristics of
their models. Horvat et al. \cite{14} extended the concept of
gravastar by introducing an electric field and examined the
stability of both the internal and external regions. In a similar
vein, Turimov et al. \cite{15} studied the effects of a magnetic
field on the gravastar geometry and obtained precise solutions for a
slowly rotating gravastar. The stability of thin-shell gravastars
developed from inner nonsingular de Sitter manifold and outer
charged noncommutative black hole solution by using Visser's cut and
paste approach is discussed in \cite{15q}.  {The
stability of gravastars solution with quintessence field is explored
in \cite{tt3} by using exact black hole solution surrounded by
quintessence field. Also, the stability of thin-shell gravastars by
using regular black holes through linearized radial perturbation is
discussed in \cite{tt4}-\cite{tt5}}.

In the field of cosmology, researchers have developed modified
gravitational theories as alternatives to general relativity (GR) in
order to explain the causes behind the expansion of the universe.
These theories introduce different approaches, such as $f(R,T)$
gravity \cite{16}, $f(R,T, R_{\alpha\beta}T^{\alpha\beta})$ gravity
\cite{17}, and $f(\mathcal{G},T)$ theory \cite{18}, where $R$
represents the curvature invariant, $T$ represents the trace of the
energy-momentum tensor (EMT), and $\mathcal{G}$ represents the
Gauss-Bonnet invariant. The study of gravastar geometries has
motivated researchers to investigate the effects of these extended
gravitational theories on different types of gravastars. In the
context of $f(R,T)$ gravity, Das et al. \cite{19} explored the
concept of gravastar geometry and analyzed its characteristics using
graphical methods for various equations of state (EoSs). Shamir and
Ahmad \cite{20} derived non-singular solutions for gravastars and
obtained mathematical expressions for various physical parameters
within the framework of $f(\mathcal{G},T)$ theory. Sharif and Waseem
\cite{21} investigated the influence of the Kuchowicz metric
potential on the structure of gravastars in $f(R,T)$ theory.  Sharif
and Naz \cite{25},\cite{26} studied important aspects of gravastars
in the absence and presence of an electric field within the
framework of $f(R,T^{2})$ gravity. Usmani et al. \cite{27} utilized
these Killing vectors to investigate various aspects of gravastars
with electric charge, obtaining solutions that corresponded to
different eras of gravastar evolution. They specifically focused on
the influence of conformal vectors on the gravastar structure.
Sharif and Waseem \cite{28} examined the impact of the charge on the
gravastar's structure within the framework of $f(R,T)$ theory,
taking into account the conformal Killing vectors. Bhar and Rej
\cite{29} proposed a charged gravastar model that incorporated
conformal motion within the framework of $f(\mathcal{T})$ theory.
They explored the consequences of this motion on the properties of
the charged gravastar. Sharif and his collaborators
\cite{30},\cite{31} investigated analytic solutions for both charged
and uncharged gravastar models using conformal Killing vectors in
the context of $f(R,T^2)$ theory.  { The study
gravastars in $f(Q,T)$ gravity is presented by Pradhan et al
\cite{gh1}.  They explored the physical characteristics and energy
constraints of the developed structure.}

 {Recently, the new gravastar model in the background of cylinderical spacetime was presented in \cite{fg1}. The model is free from singularities and information paradoxes, making it stable and physically viable. In the framework of Rastall gravity, the impact of charge on the possible formation of isotropic spherically symmetric gravastar configuration is explored in \cite{fg2}. Rosa et al. \cite{fg3} examined the observable characteristics of thin-shell gravastars in two astrophysical models: one in which they are orbited by hot-spots and the other in which they are encircled by optically-thin accretion disks. They employed numerical reverse ray-tracing techniques to generate the attendant observables \cite{fg3}. Regarding the accretion disk data, results showed that smooth gravastar configurations cannot recreate shadow measurements when internal emission is assumed, since there is no substantial gravitational redshift impact \cite{fg3}.}

 {Researchers are interested in investigating the
physical characteristics of the wormhole structure in different
modified theories of gravity and also evaluating the stability of
the shell around the wormhole structure. In the context of Rastall
gravity, the WH solutions are provided in \cite{ta10} by the
utilization of the phantom regime with conformal symmetry. The
stability of the thin-shell around the computed WH solutions was
investigated in \cite{ta11} after they investigated the novel WH
solutions against the backdrop of teleparallel gravity. In $f(R,T)$
gravity, the research of WH solution and thin-shell surrounding the
WH geometry is presented in \cite{ta12}. The stability of the
thin-shell around the WH geometry and the investigation of
traversable WH solutions in $f(Q)$ that exhibit twin peak
quasi-periodic oscillations are detailed in \cite{ta13}. The
consequences of quantum wave dark matter, the stability of
thin-shell around the estimated WH geometry, and the WH solutions in
the GR framework were examined in \cite{ta14}. The study of compact
stars in different modified theories of gravity is presented in
\cite{jimr1}-\cite{jimr14}.}

The presence of dark energy in black holes aligns with the standard
model of cosmology, which indicates its dominance in the universe
\cite{38q},\cite{39q}. This interaction between dark energy and
black holes, similar to the cosmological constant or vacuum energy,
can have significant implications \cite{41q}. Astronomical
observations have shown that the universe is expanding at an
accelerating rate, implying the existence of negative pressure,
which can be explained by quintessence dark energy
\cite{42q}-\cite{45q}.  Theoretical advancements propose that
one-dimensional strings, rather than point-like particles, are the
fundamental units of nature. Exploring Einstein's equations with
string clouds is crucial, as relativistic strings can be used to
construct appropriate models \cite{54q}. The concept of a cloud of
strings as the source of the gravitational field was initially
introduced in \cite{54q}. Lately, who discovered an exact solution
for a Schwarzschild black hole surrounded by strings
\cite{54q},\cite{55q}. The effects of both quintessence and cloud of
strings on the black hole thermodynamics are explored in \cite{60q}.
 {Further, the investigation of null geodesics of
Schwarzschild black hole in the framework of quintessence and the
cloud of strings are explored in \cite{tt1}.  The impact of the
cloud of string and quintessence field on the stability of higher
dimensional thin-shell wormholes are studied in \cite{tt2}.}

The concept of gravastars, which are theoretical alternatives to
black holes, has captured the interest of scientists and researchers
in the field of astrophysics. However, there is still much to be
explored and understood about these intriguing objects. The
motivation behind this paper lies in the desire to further
investigate the exact solution of gravastars within the framework of
general relativity, specifically by incorporating the cloud of
strings and quintessence. The paper is organized as follows: In
section-II, we present the Einstein field equations with the
presence of a modified matter source as the cloud of strings and
quintessence. In section-III, we calculate the exact solutions of
the inner, and outer regions of gravastars. Then, we consider the
exterior region as a black hole surrounded by a cloud of strings and
quintessence. Then, we match these spacetimes through well-known
Darmoise-Israel formalism as present in In section-IV. Section-V is
devoted to exploring the stability of gravastars using linearized
radial perturbation and also calculating the proper length, energy
content, and entropy of the shell. The last section is devoted to
presenting the concluding remarks. Throughout the calculations, we consider the geometric
units $G = c = 1$, and the spacetime signature $(+, -, -, -)$.

\section{Assessing the Impacts of cloud of strings and quintessence on Einstein's Field Equations}\label{sec2}

The focus of this section is on the 4-dimensional spacetime that is
both spherically symmetric and static, and it is bounded by a
spherical surface. Using the Schwarzschild coordinates, the
respective metric is presented as follows
\begin{eqnarray}
ds^2 = e^{\epsilon(r)}dt^{2} -
e^{\varepsilon(r)}dr^{2}-r^{2}\sin^{2}\theta d
\phi^{2}-r^2d\theta^{2}, \label{1}
\end{eqnarray}
here the gravitational functions of temporal and radial coordinates
are denoted by $\epsilon(r)$ and $\varepsilon(r)$, respectively. By
using the modified form of matter, the respective Einstein field
equations for the metric (\ref{1}) become
\begin{eqnarray}
 &&   G_{ij}=R_{ij}-\frac{1}{2}\,g_{ij}\,R= T^{\text{eff}}_{ij}, \quad\quad i,j=0,1,2,3,\label{eq2}
\end{eqnarray}
where
\begin{eqnarray}
T^{\text{eff}}_{ij}= \Theta_{ij}+\hat{T}_{ij}+\hat{\Theta}_{ij},
\label{eq3}
\end{eqnarray}
here $\Theta_{ij}$ represents the matter due to cloud of string and
$\hat{\Theta}_{ij}$ depicts the matter influenced by quintessence
fields. Consequently, in the background of strings of clouds, the
respective Lagrangian density can be written as \cite{122}
\begin{equation}\label{S1}
L_s=-\frac{k}{2}\Sigma^{ij}\Sigma_{ij},
\end{equation}
where the tension of the string and bi-vector is denoted by a
constant $k$. In this regards, we obtain the following relation
\begin{equation}\label{S3}
\Sigma^{ij}=\epsilon^{a\beta}\frac{\partial x^i}{\partial
\lambda^{a}}\frac{\partial x^j}{\partial \lambda^{\beta}},
\end{equation}
here relation for the parameterization of the world sheet is
referred as $\lambda^{a} (\lambda^{a}=\lambda^0, \lambda^1)$ and
Levi-Civita tensor is represented by  $\epsilon^{a \beta}$.  By
using induced metric, it can be described for the string as follows
\cite{122}
\begin{equation}\label{S4}
h_{a \beta}=g_{ij}\frac{\partial x^i}{\partial
\lambda^{a}}\frac{\partial x^j}{\partial \lambda^{\beta}}.
\end{equation}
Consequently, the some important identities is obtained from
$\Sigma^{i j}$ given as  \cite{122}, \cite{123}
\begin{equation}\label{S5}
\Sigma^{i[a} \Sigma^{\beta \sigma]}=0, \quad \Sigma^{i a} \Sigma_{a
\sigma} \Sigma^{\sigma j} =\textbf{h} \Sigma^{ji}, \quad \nabla_i
\Sigma^{i[ a} \Sigma^{\beta \sigma]}=0,
\end{equation}
where the determinant of $h_{a \beta}$ is denoted with $\textbf{h}$.

Further, by varying the Lagrangian density concerning the metric
tensor $g_{ij}$, we get  \cite{123}
\begin{equation}\label{S6}
\Theta_{ij}=\rho_s\frac{\Sigma^{i
a}\Sigma^{j}_{a}}{\sqrt{-\textbf{h}}},
\end{equation}
where $\rho_s$ denotes the density of the string cloud.  We obtain
the following expression $\partial_i(\sqrt{-g}\Sigma^{i a})=0$ by
considering the identities mentioned in Eq.(\ref{S5}). In the
background of string clouds, the respective components of the
stress-energy tensor become~ \cite{124}:
\begin{equation}\label{11}
\Theta_{tt}=\Theta_{rr}=-\frac{a}{r^2}\;\;\;\;\;\;\;\Theta_{\theta\theta}=\Theta_{\phi\phi}=0,
\end{equation}
where the parameter $a$ denotes the cloud of strings. Also, we
obtain the following relation for quintessence matter distribution
as follows~ \cite{125}
\begin{equation}\label{q}
L_q=-\frac{1}{2}g^{ij}\partial_i\Psi \partial_j\Psi-V(\Psi).
\end{equation}
The stress-energy-momentum tensor components that are physically
viable under the influence of the quintessence field ($\Psi$) is
specified by $V(\Psi)$, which denotes the potential term of the
quintessence field.~ \cite{122}, \cite{126}
\begin{equation}\label{q1}
\hat{\Theta}_{tt}=\hat{\Theta}_{rr}=\rho_{q}\;\;\;\;
\hat{\Theta}_{\theta\theta}=\hat{\Theta}_{\phi\phi}=-\frac{\rho_{q}}{2}(3\omega_q+1),
\end{equation}
where $\omega_q$ and $\rho_q$ denote the quintessence field
parameter and quintessence density, respectively, and are used to
characterize the system's internal makeup in the presence of the
matter source $\hat{T}_{ij}$. To analyze the physical composition of
spacetime, we examine the energy density of an isotropic matter
distribution that covers the area. Here, the radial pressure is
represented by $p$ and the energy density is depicted with $\rho$.
Further, the respective components of stress-energy tensor
$\hat{T}_{ij}$ become
\begin{eqnarray}
&&
\hat{T}_{tt}=\rho,~~~\hat{T}_{rr}=-p_r,~~~~\hat{T}_{\theta\theta}=\hat{T}_{\phi\phi}=-p_t.
\label{eq7}
\end{eqnarray}
Consequently, we get the modified form of field equations
(\ref{eq2}) in the presence of a cloud of strings and quintessence
field given as
\begin{eqnarray}
\label{eq8} && \rho +\rho_{q}+\frac{a
}{r^2}=\frac{1}{8\pi}\bigg[\frac{1}{r^2}-e^{-\varepsilon(r)}\left(\frac{1}{r^2}-\frac{\varepsilon^{\prime}(r)}{r}\right)\bigg],
\\\label{eq9}
&& {p}-\rho_{q}-\frac{a
}{r^2}=\frac{1}{8\pi}\bigg[e^{-\varepsilon(r)}\left(\frac{1}{r^2}+\frac{\epsilon^{\prime}(r)}{r}\right)-\frac{1}{r^2}\bigg],~~~~~
\\\label{eq10} && {p}+\frac{1}{2} \rho_{q} (3
w_q+1)=\frac{1}{8\pi}\bigg[\frac{e^{-\varepsilon(r)}}{4}\Big(2\epsilon^{\prime\prime}(r)+\epsilon^{\prime2}(r)-\varepsilon^{\prime}(r)\epsilon^{\prime}(r)+2\frac{\epsilon^{\prime}(r)-\varepsilon^{\prime}(r)}{r}\Big)\bigg],~~~
\end{eqnarray}
where the derivative with respect to radial component $r$ is denoted
by primes. Now, by  solving the above field equations for $\rho$,
$\rho_{q}$, and $p$, we get the following expressions for the
corresponding quantities:
\begin{eqnarray}
\label{eq11}
 \rho&&=-\frac{(a -1) (3 \omega_q+1)}{24 \pi  r^2 (\omega_q+1)}+\frac{e^{-\varepsilon (r)} \left(-\left(r^2 \epsilon ''(r)-r (3 \omega_q+4) \varepsilon '(r)+3 \omega_q+1\right)\right)}{24 \pi  r^2 (\omega_q+1)}+\frac{e^{-\varepsilon (r)} \left(r \varepsilon '(r)+2\right) \epsilon '(r)}{48 \pi  r \omega_q+48 \pi  r}\nonumber\\&&-\frac{e^{-\varepsilon (r)} \epsilon '(r)^2}{48 \pi  (\omega_q+1)},
\\\label{eq12}
 {p}&&=\frac{2 a +6 a  \omega_q-6 \omega_q-2}{48 \pi  r^2 (\omega_q+1)}+\frac{e^{-\varepsilon (r)} \left(2 r^2 \epsilon ''(r)-2 r \varepsilon '(r)+6 \omega_q+2\right)}{48 \pi  r^2 (\omega_q+1)}+\frac{e^{-\varepsilon (r)} \epsilon '(r) \left(r^2 \left(-\varepsilon '(r)\right)+6 r \omega_q+4 r\right)}{48 \pi  r^2 (\omega_q+1)}\nonumber\\&&+\frac{e^{-\varepsilon (r)} \epsilon '(r)^2}{48 \pi  (\omega_q+1)}, \\\label{eq13}
 \rho_{q} &&=\frac{4-4 a }{6 r^2 (\omega_q+1)}+\frac{e^{-\varepsilon (r)} \left(2 r^2 \epsilon ''(r)-2 r \varepsilon '(r)-4\right)}{6 r^2 (\omega_q+1)}+\frac{e^{-\varepsilon (r)} \left(r^2 \left(-\varepsilon '(r)\right)-2 r\right) \epsilon '(r)}{6 r^2 (\omega_q+1)}+\frac{e^{-\varepsilon (r)} \epsilon '(r)^2}{6 (\omega_q+1)}.
\end{eqnarray}
Now, for the GR gravity the energy conservation equation under the
current scenario can be provided as
\begin{equation}\label{eq14}
\frac{\epsilon '}{2}(\rho+p)+\frac{dp}{dr}=0.
\end{equation}
From the equation mentioned above, it becomes apparent that for a
gravitating system to be in a state of equilibrium, the force of
gravity must be counter balanced by an equal pressure gradient known
as the hydro-static force.

Further, we are interested in developing gravastar structure in the
background of the cloud of strings and quintessence by using the
exact field equations with a modified matter source. Overall, the
gravastar configuration with the cloud of strings and quintessence
is a fascinating model that may shed light on the nature of dark
energy and the origins of the universe \cite{54q,55q}. Gravastar is
a hypothetical model that is formed from the collapse of matter in
the presence of a negative pressure source which can be better
explained by quintessence field. Quintessence is a form of dark
energy that is believed to be responsible for the acceleration of
the universe's expansion \cite{42q}-\cite{45q}. In this model, the
gravastar consists of a thin-shell of negative pressure, surrounded
by a thick shell of ultra-relativistic cosmic string. The cosmic
string is a hypothetical one-dimensional object that is formed from
the stretching of space-time due to the presence of a concentrated
mass.  {In the gravastar structure, the quintessence
shell is supported by the cosmic string, which generates the
required self-gravity. The system's self-gravity is enhanced by the
quintessence's negative pressure, which results in a stable
equilibrium. The ability for stable orbits to occur inside the
object is an intriguing feature of the gravastar structure. This is
because the cosmic string shell generates a repulsive force, while
the quintessence shell pulls on neighboring masses gravitationally.
The combination of these forces can lead to the creation of a stable
region within the gravastar where orbits can be maintained.}

\section{Gravastar configuration with string and quintessence}

In this section, we develop the gravastar structure by using the
field equation in the framework of a modified matter source. For
this purpose, we develop the gravastar by understanding its
geometrical structure which can be partitioned into three different
regions. The behavior of matter contents of these regions can be
characterized through a specific type of EoS. Such geometrical
structure is partitioned into interior ($0\leq r<r_1$), intermediate
or thin-shell ($r_1< r<r_2$) and exterior region ($r_2<r$). Here,
$r_1$ and $r_2$ denote the radius of inner and outer regions. Also,
the thickness of the intermediate region is referred to as
$r_2-r_1$. The specific EoS for these regions is expressed as
\begin{itemize}
\item $p=-\sigma$ for inner region;
\item $p=\sigma$ for intermediate region;
\item $p=0=\sigma$ for the outer region.
\end{itemize}
In the following subsections, we discuss in detail the solution of
these regions by using the respective EoS.

\subsection{Interior region of the gravastar with string and quintessence}

The relationship between the metric potentials and the physical
parameters like $\rho$ and $p$ from Eq. (\ref{eq11}) and Eq.
(\ref{eq12}) by using the interior area as stated in Mazur,
Mottola's work is provided below:
\begin{eqnarray}\label{eq15}
p=-\rho,
\end{eqnarray}
the parameterized version of this EoS is known as the dark energy
EoS with $w=-1$. This negative pressure acts radially outwards from
the center of the spherically symmetric gravitating system,
counteracting the inward gravitational attraction of the shell. It
can be shown, by using Eq. (\ref{eq15}) in Eq. (\ref{eq14}) as
\begin{eqnarray}\label{eq16}
p=-\rho=-\rho_{c},
\end{eqnarray}
where $\rho_{c}$ represents the critical density of the gravastar.
Now, with the help of Eq. (\ref{eq15}) within the scope of the Eqs.
(\ref{eq11}) and (\ref{eq12}), we have following relation
\begin{eqnarray}\label{eq17}
\frac{e^{-\varepsilon(r)}
\left(\epsilon'(r)+\varepsilon'(r)\right)}{8 \pi  r}=0.
\end{eqnarray}
Now, by solving the Eq. (\ref{eq17}), we have the following
connection between gravitational metric components:
\begin{eqnarray}\label{eq18}
\epsilon(r)=-\varepsilon(r)+C_{1},
\end{eqnarray}
here $C_{1}$ represents the constant of integration. Now, by
plugging the Eqs. (\ref{eq16}) and (\ref{eq18}) in Eq. (\ref{eq11}),
we have following differential equation:
\begin{eqnarray}\label{eq19}
\rho_{c}=\frac{e^{-\varepsilon(r)} \left(r \left(r
\varepsilon''(r)+\varepsilon'(r) \left(-r \varepsilon'(r)+3
\omega_q+3\right)\right)-(3 \omega_q+1) \left((a -1)
e^{\varepsilon(r)}+1\right)\right)}{24 \pi  r^2 (\omega_q+1)}.
\end{eqnarray}
It is noticed that the above equation is depending on one
gravitational metric components, i.e., $\varepsilon(r)$, and
quintessence field $\omega_q$. Now, we can solve the above
differential equation for the specific value of $\omega_q$, which
should posing quintessence matter. In this regard, we shall choose,
$\omega_q=\frac{-2}{3}$ (quintessence matter), and get the following
expression for $\varepsilon(r)$ as:
\begin{eqnarray}\label{eq20}
\varepsilon(r)=-\log \left(-a +\frac{C_{2}
r}{2}-\frac{C_{3}}{r}-\frac{1}{3} 8 \pi  r^2 \rho_{c} +1\right),
\end{eqnarray}
where $C_{2}$ and $C_{3}$ are constants of integration. For the
purpose of the regular solution at the center, we must impose the
condition, i.e., $C_{3}=0$. Finally, we have the following
simplified relation for $g_{rr}$ metric component written as
\begin{eqnarray}\label{eq20}
\varepsilon(r)=-\log \left(-a +\frac{C_{2} r}{2}-\frac{1}{3} 8 \pi
r^2 \rho_{c}+1\right).
\end{eqnarray}
Now, by using the Eq. (\ref{eq20}) in Eq. (\ref{eq18}), we have the
following $g_{tt}$ metric component:
\begin{eqnarray}\label{eq21}
\epsilon(r)=\log \left(-a +\frac{C_{2} r}{2}-\frac{1}{3} 8 \pi  r^2
\rho_{c}+1\right)+C_{1}
\end{eqnarray}

\subsection{Intermediate region}

Herein, we calculate the exact solution for intermediate region with
physical parameters in the framework of EoS, which is defined as
\begin{eqnarray}\label{eq22}
p=\rho.
\end{eqnarray}
Now, by plugging the Eqs. (\ref{eq11}) and (\ref{eq12}) in Eq.
(\ref{eq22}), we have the following relation
\begin{eqnarray}\label{eq23}
p-\rho&&= \frac{e^{-\varepsilon(r)}}{24 \pi  r^2
(\omega_q+1)}\left[r \left(2 r \epsilon''(r)+\epsilon'(r) \left(-r
\varepsilon'(r)+3 \omega_q+1\right)+r \epsilon'(r)^2-(3 \omega_q+5)
\varepsilon'(r)\right)\nonumber\right.\\&&\left.+2 (3 \omega_q+1)
\left((a -1) e^{\varepsilon(r)}+1\right)\right].
\end{eqnarray}
The intermediate shell is assumed to be created by an
ultra-relativistic fluid with a non-vacuum background and an EoS
$p=-\rho$. In the non-vacuum zone, it is difficult to determine the
solution to the complicated set of field equations. To avoid this
inquiry, we will make some approximations and discover an analytical
answer, namely, $0<e^{\varepsilon(r)}<<1$. We use the embedded
technique to find the precise answer, i.e. the product terms of
metric functions and  derivative of metric component become vanish
as  $e^{\varepsilon(r)}\epsilon'(r)\Rightarrow0$. In this regard, we
get a modified version of field equations in the simplified form as:
\begin{eqnarray}\label{eq24}
\frac{e^{-\varepsilon(r)} \left(2 (3 \omega_q+1) \left((a -1)
e^{\varepsilon(r)}+1\right)-r (3 \omega_q+5)
\varepsilon'(r)\right)}{24 \pi  r^2 (\omega_q+1)}=0.
\end{eqnarray}
On solving the above equation, one can get the following expression
for $g_{rr}$ metric component:
\begin{eqnarray}\label{eq25}
\varepsilon(r)=-\log \left(-a +3^{2/3} e^{-C_{4}} r^{2/3}+1\right),
\end{eqnarray}
with $C_{4}$ as integrating constant. For the $g_{rr}$ metric
component, we make a use of Eq. (\ref{eq14}) and get the following
relation of the metric components:
\begin{eqnarray}\label{eq26}
&&\frac{e^{-\varepsilon(r)}}{48 \pi  r^3 (\omega_q+1)}\left[r
\left(-3 \varepsilon'(r) \left(r^2 \epsilon''(r)+2 \omega_q\right)+2
r \epsilon'(r)^2 \left(-r \varepsilon'(r)+3 \omega_q+2\right)+r^2
\epsilon'(r)^3+2 r \left(r
\epsilon^{(3)}(r)\nonumber\right.\right.\right.\\&&\left.\left.\left.+(3
\omega_q+2) \epsilon''(r)-\varepsilon''(r)\right)+\epsilon'(r)
\left(r \left(r \left(4 \epsilon''(r)-\varepsilon''(r)\right)-6
(\omega_q+1) \varepsilon'(r)+r
\varepsilon'(r)^2\right)\nonumber\right.\right.\right.\\&&\left.\left.\left.+2
(a -1) (3 \omega_q+1) e^{\varepsilon(r)}-2\right)+2 r
\varepsilon'(r)^2\right)-4 (3 \omega_q+1) \left((a -1)
e^{\varepsilon(r)}+1\right)\right]=0.
\end{eqnarray}
Now, again we are adopting the embedded procedure as mentioned above
and get the following final differential equation:
\begin{eqnarray}\label{eq27}
\frac{(3 \omega_q+1) e^{-\varepsilon(r)} \left((a -1)
e^{\varepsilon(r)} \left(r \epsilon'(r)-2\right)-2\right)}{24 \pi
r^3 (\omega_q+1)}=0.
\end{eqnarray}
By plugging the Eq. (\ref{eq25}) in Eq. (\ref{eq27}), one can get
the following expression for $g_{tt}$ metric component:
\begin{eqnarray}\label{eq28}
\epsilon(r)=\frac{3\times 3^{2/3} e^{-C_{3}} r^{2/3}}{a -1}+C_5,
\end{eqnarray}
where $C_5$ is an integrating constant. By using these metric
function in the energy conservation equation for stiff matter
distributions, we get
\begin{eqnarray}\label{eq28a}
p=\rho=C_6 e^{-\frac{3\ 3^{2/3} e^{-C_3} r^{2/3}}{a -1}},
\end{eqnarray}
with integrating constant $C_6$.

\subsection{The Exterior Region}

A black hole with quintessence and cloud of strings would be a very
interesting and complex phenomenon. Quintessence is a theoretical
form of dark energy that is thought to be responsible for the
accelerating expansion of the universe. It is a hypothetical scalar
field that permeates all of space and time, and it has the potential
to affect the behavior of objects near a black hole. Cloud strings
are another hypothetical phenomenon that could potentially exist in
the vicinity of a black hole \cite{54q,55q}. They are thin, long,
and extremely dense configurations of matter that are thought to
form in the early universe. These cloud strings could exist in the
vicinity of a black hole and modify its properties. Combining these
two phenomena, we can imagine a black hole that is surrounded by a
cloud of quintessence and cloud strings. This could have a number of
interesting effects on the black hole, including altering its mass,
spin, and other properties. For example, the presence of the
quintessence field could have the effect of slowing down the rate at
which matter and energy fall into the black hole
\cite{42q}-\cite{45q}. This could in turn reduce the rate at which
the black hole grows in mass over time. The presence of the cloud
strings could also modify the behavior of particles in the vicinity
of the black hole, potentially leading to novel and unexpected
phenomena. In the present work, we are interested to observe the
affects of both cloud of strings and quintessence on the geometrical
structure of gravastars. In this regards, we consider the exact
solution of the black hole geometry surrounded by cloud of strings
and quintessence as an exterior manifold \cite{60q}. The line
element of such black hole geometry can be expressed as  \cite{60q}
\begin{eqnarray}\label{28}
ds^2=\left(-a-\frac{2 m}{y}-\frac{\gamma }{y^{3 \omega
_q+1}}+1\right)dt^2-\left(-a-\frac{2 m}{y}-\frac{\gamma }{y^{3
\omega _q+1}}+1\right)^{-1}dr^2-r^2(d\theta^2+\sin\theta^2d\phi^2).
\end{eqnarray}
where $\gamma$ stands for the Kiselev parameter. In the absence of
$\gamma$ and $a$, the considered black hole geometry reduces to the
Schwarzschild black hole.

In the following section, we match the interior and exterior regions
through well-known cut and paste approach \cite{q1,q2,q3}. Then, we
explore the effects of cloud of strings and quintessence field on
the stability as well as physical characteristics of the developed
gravastar structure.

\section{Matching of interior and exterior regions through junction conditions}

In order to develop the geometry of gravastar, we match the exact
solutions of exterior (+) and interior (-) regions through junction
conditions. These geoemtries are matched at the hypersurface by
implementing Visser cut and paste approach \cite{q1,q2,q3}. Hence,
we cut inner and outer manifolds into the following regions as
\begin{eqnarray}\label{3aa}
\mathcal{M}^{\pm}=\left\lbrace r^{\pm}\leq y,y>r_{h}\right\rbrace,
\end{eqnarray}
during this Darmois-Israel formalism, it is necessary to maintain
the radius of shell ($y$) must be greater then the radius of event
horizon of the considered exterior black hole spacetime. In this
approach, the developed structure is free from event horizon as well
as the spacetime singularity \cite{q1,q2,q3}. Both, spacetimes are
connected at the hypersurface ($\triangle$) which is
(2+1)-dimensional manifold given as
\begin{eqnarray}\label{4aa}
\triangle=\left\lbrace r^{\pm}=y,y>r_{h}\right\rbrace.
\end{eqnarray}

Hence, we get the unique regular spacetime which can be written as
mathematically as  $\mathcal{M}=\mathcal{M}^{-}\cup
\mathcal{M}^{+}$. The coordinates of hypersurface and manifolds have
the  following form $\eta^{i}=(\tau,\theta,\phi)$,\quad
$J^{\gamma}_\pm=(t_\pm,r_\pm,\theta_\pm,\phi_\pm)$, respectively.
Here proper time is represented with $\tau$. Also, the coordinates
of hypersurface and the manifolds are connected through the
following coordinate transformation
\begin{eqnarray}\label{5aa}
g_{ij}=\frac{\partial J^{\gamma}}{\partial\eta^{i}}\frac{\partial
J^{\beta}}{\partial\eta^{j}}g_{\gamma\beta}.
\end{eqnarray}
Consequently, the hypersurface parametric equation can be written as
\begin{eqnarray}\nonumber
\triangle:R(r,\tau)=r-y(\tau)=0.
\end{eqnarray}
The thin layer of matter content located at the shell is very
important to discuss the dynamics as well as stable configuration of
the shell. The surface pressure and the energy density of the matter
can be evaluated through the Lanczos equations given as
\cite{q1,q2,q3}
\begin{equation}\label{6aa}
S_{\beta}^{\alpha}=\frac{1}{8\pi}(\delta_{\beta}^{\alpha}
\zeta_{\gamma}^{\gamma}-\zeta_{\beta}^{\alpha}),
\end{equation}
where $\zeta_{\alpha\beta}=K^{+}_{\alpha\beta}-K^{-}_{\alpha\beta}$
and $K^{-}_{\alpha\beta}$ represents the components of extrinsic
curvature. The energy-momentum tensor for the perfect fluid can be
written as ${S^{\alpha}}_{\beta}=diag(\vartheta,\zeta,\zeta)$ where
$\vartheta$ and $\zeta$ denotes the energy density and surface
pressure of ideal fluid.  Also, the components of extrinsic
curvature are written  as
\begin{equation}\label{7aa}
{K_{\alpha\beta}^{\pm}}= -n_{\mu}^\pm \left[\frac{\partial^2
J^{\mu}_\pm}{\partial \eta^{\alpha}
\eta^{\beta}}+\Gamma^{\mu}_{\lambda\nu}\left(\frac{\partial
J^{\lambda}_\pm}{\partial\eta^{\alpha}}\right)\left(\frac{\partial
J^{\nu}_\pm}{\partial\eta^{\beta}}\right)\right],
\end{equation}
where $n_{\mu}$ denoted the normal vector which can be given as
\begin{eqnarray}\label{7}
n_{\pm}^{\mu}&=&\begin{cases}(\dot{y}(1-a-\frac{1}{3} 8 \pi  y^2
\rho _c+\frac{\text{C}_2 y}{2})^{-1},\sqrt{1-a-\frac{1}{3} 8 \pi y^2
\rho _c+\frac{\text{C}_2 y}{2}
+\dot{y}^2},0,0),\quad \text{for interior region}\\
(\dot{y}(-a-\frac{2 m}{y}-\frac{\gamma }{y^{3 \omega
_q+1}}+1)^{-1},\sqrt{-a-\frac{2 m}{y}-\frac{\gamma }{y^{3 \omega
_q+1}}+1 +\dot{y}^2},0,0),\quad \text{for exterior
region}\end{cases},
\end{eqnarray}
here over dot represents the derivative with respect to the proper
time. By employing  the Lanczos equations, we obtain
\begin{eqnarray}\label{10qa}
4 \pi  y\vartheta&=&-\sqrt{-a-\frac{2 m}{y}-\frac{\gamma }{y^{3
\omega _q+1}}+1+\dot{y}^2}+\sqrt{-a-\frac{1}{3} 8 \pi  y^2 \rho
_c+\frac{\text{C}_2 y}{2}+1+\dot{y}^2},
\\\label{10aa}
8 \pi  y\zeta&=&\frac{3 \left(4 a-4 \dot{y}^2-3 \text{C}_2 y-4
\ddot{y} y-4\right)+64 \pi  y^2 \rho _c}{\sqrt{6} \sqrt{-6 a+6
\dot{y}^2-16 \pi y^2 \rho _c+3 \text{C}_2 y+6}}+\frac{ \left(2 y^{3
\omega _q} \left(y \left(-a+\dot{y}^2+\ddot{y}
y+1\right)-m\right)-\gamma +3 \gamma \omega _q\right)}{y^{3 \omega
_q+1}\sqrt{-a+\dot{y}^2-\frac{2 m+\gamma y^{-3 \omega _q}}{y}+1}}.
\end{eqnarray}

The surface pressure and energy density of shell follows the energy
conservation constraints as
\begin{equation}\nonumber
\zeta\frac{d}{d\tau}(4\pi y^2)+\frac{d}{d\tau}(4\pi y^2\vartheta)=0,
\end{equation}
which leads to
\begin{equation}\nonumber
\vartheta'=-\frac{2(\vartheta+\zeta(\vartheta))}{y}.
\end{equation}

It is noted that the equilibrium position of the shell ($y_0$) leads
to vanishing of derivative of shell radius with respect to the
proper time, i.e., $\dot{y_0}=0=\ddot{y_0}$.. At equilibrium shell
radius, shell does not move in the radial direction. Hence, we have
\begin{eqnarray}\label{9aa}
4 \pi  y_0\vartheta_0&=&-\sqrt{-a-\frac{2 m}{y_0}-\frac{\gamma
}{y^{3 \omega _q+1}_0}+1}+\sqrt{-a-\frac{1}{3} 8 \pi  y^2_0 \rho
_c+\frac{\text{C}_2 y_0}{2}+1},
\\\label{10aa}
8 \pi  y_0\zeta_0&=&\frac{12 a+64 \pi  y^2_0 \rho _c-9 \text{C}_2
y_0-12}{\sqrt{6} \sqrt{-6 a-16 \pi  y^2_0 \rho _c+3 \text{C}_2
y_0+6}}+\frac{y_0^{-3 \omega _q-1} \left(-2 ((a-1) y_0+m) y_0^{3
\omega _q}-\gamma +3 \gamma  \omega _q\right)}{\sqrt{-a-\frac{2
m+\gamma y_0^{-3 \omega _q}}{y_0}+1}},
\end{eqnarray}
where  $\vartheta_0$ and $\zeta_0$ are the energy density and
pressure at equilibrium position, respectively.

\section{Some Physical characteristics of Gravastars}

This section is devoted to examining the impact of the cloud of
strings and quintessence field on different physical features of
gravastar. In this context, we shall calculate the stability through
linearized perturbation. Then, we observe the proper length, shell
energy and entropy of the gravastars.

\subsection{Stability with Linearized Radial Perturbation}

First, we calculate the equation of motion of the shell from Eq.
(\ref{10qa}) as
\begin{equation}
\dot {y}^2+\Omega(y)=0\,,
\end{equation}
where $\Omega(y)$ depicts the potential function which can be
calculated as
\begin{eqnarray}\nonumber
\Omega(y)&=& -a+\frac{\rho _c \left(-8 \pi  y^3 \rho _c+3 \text{C}_2
y^2+6 \gamma  y^{-3 \omega _q}\right)}{72 \pi  \vartheta ^2
y}+\frac{m \rho _c}{6 \pi  \vartheta ^2 y}-\frac{4}{3} \pi  y^2
\left(\rho _c+3 \pi  \vartheta ^2\right)-\frac{\gamma  y^{-6 \omega
_q-4} \left(\gamma +\left(\text{C}_2 y^2+4 m\right) y^{3 \omega
_q}\right)}{64 \pi ^2 \vartheta
^2}\\\label{9aa}&-&\frac{\left(\text{C}_2 y^2+4 m\right)^2}{256 \pi
^2 \vartheta ^2 y^4}+\frac{\text{C}_2 y}{4}-\frac{m}{y}-\frac{1}{2}
\gamma  y^{-3 \omega _q-1}+1.\label{23a}
\end{eqnarray}
We are interested in perturbing the system about equilibrium shell
radius to explore the stability of the gravastars. In this regard,
we expand potential function around $y_0$ using Taylor series up to
2nd order terms $(y-y_{0})$. Hence, we get
\begin{widetext}
\begin{eqnarray}
\Omega(y) &=&  \Omega(y_0) + \Omega^\prime(y_0) ( y - y_0) +
\frac{1}{2} \Omega^{\prime\prime}(y_0) ( y - y_0)^2  +
\mathcal{O}\left[( y - y_0)^3\right]\,.\label{24}
\end{eqnarray}
\end{widetext}
By using the conservation equation, we get
\begin{eqnarray}\nonumber
\Omega^{\prime}(y)&=& \frac{y^{-6 \omega _q-5}}{1152 \pi ^2
\vartheta (y)^3} \left(-2 (\zeta(y)+\vartheta (y)) \left(y^{3 \omega
_q} \left(-16 \pi  y^3 \rho _c+3 \text{C}_2 y^2+12 m\right)+6 \gamma
\right){}^2+96 \pi ^2 \vartheta (y)^3 y^{3 \omega _q+3}
\right.\\\nonumber&\times&\left.\left(y^{3 \omega _q} \left(-32 \pi
y^3 \rho _c+3 \text{C}_2 y^2+12 m\right)+6 \gamma \left(3 \omega
_q+1\right)\right)+2 \vartheta (y) \left(4 y^{3 \omega _q} \left(2
\pi y^3 \rho _c+3 m\right)+3 \gamma  \left(3 \omega
_q+2\right)\right) \right.\\\nonumber&\times&\left.\left(y^{3 \omega
_q} \left(-16 \pi y^3 \rho _c+3 \text{C}_2 y^2+12 m\right)+6 \gamma
\right)+18432 \pi ^4 \vartheta (y)^4 (\zeta(y)+\vartheta (y)) y^{6
\omega _q+6}-9216 \pi ^4 \vartheta (y)^5 y^{6 \omega _q+6}\right).
\end{eqnarray}

Now, we consider the following parameter
$\eta(\vartheta)=d\zeta/d\vartheta=\zeta^{\prime}/\vartheta^\prime$
for the second derivative of the potential function as
\begin{eqnarray}\nonumber
\Omega^{\prime\prime}(y)&=& \frac{y^{-6 \left(\omega
_q+1\right)}}{576 \pi ^2 \vartheta (y)^4} \left(-3 \gamma  y^{3
\omega _q} \left(-4 \zeta(y) \vartheta (y) \left(3 \left(6 \omega _q
\left(\text{C}_2 y^2+4 m\right)+\text{C}_2 (2 \eta -5) y^2+(8 \eta
-4) m\right)-16 \pi  y^3 \rho _c \left(2 \eta
\right.\right.\right.\right.\\\nonumber&+&\left.\left.\left.\left.6
\omega _q-7\right)\right)+24 \zeta(y)^2 \left(-16 \pi  y^3 \rho _c+3
\text{C}_2 y^2+12 m\right)+\vartheta (y)^2 \left(16 \pi  y^3 \rho _c
\left(8 \eta +3 \omega _q \left(5-3 \omega
_q\right)-6\right)\right.\right.\right.\\\nonumber&+&\left.\left.\left.9
\omega _q \left(3 \omega _q \left(\text{C}_2 y^2+4 m\right)-3
\text{C}_2 y^2+4 m\right)+6 \text{C}_2 (1-4 \eta ) y^2-96 \eta
m\right)+96 \pi ^2 y^3 \left(9 \omega _q \left(\omega
_q+1\right)+2\right) \vartheta
(y)^4\right)\right.\\\nonumber&-&\left.y^{6 \omega _q}
\left(-\zeta(y) \vartheta (y) \left(-16 \pi y^3 \rho _c+3 \text{C}_2
y^2+12 m\right) \left(16 \pi  (13-2 \eta ) y^3 \rho _c+3
\left(\text{C}_2 (2 \eta -9) y^2+(8 \eta -4)
m\right)\right)\right.\right.\\\nonumber&+&\left.\left.6 \zeta(y)^2
\left(-16 \pi  y^3 \rho _c+3 \text{C}_2 y^2+12
m\right){}^2+\vartheta (y)^2 \left(32 \pi y^3 \rho _c \left(4 \pi
(15-4 \eta ) y^3 \rho _c+3 \left(\text{C}_2 (2 \eta -5) y^2+(8 \eta
-6)
m\right)\right)\right.\right.\right.\\\nonumber&-&\left.\left.\left.18
\eta \left(\text{C}_2 y^2+4 m\right)^2+9 \text{C}_2 y^2 \left(3
\text{C}_2 y^2+4 m\right)\right)+384 \pi ^2 y^3 \vartheta (y)^4
\left(4 \pi  y^3 \left(\rho _c+12 \pi  \zeta(y)^2\right)+3
m\right)\right.\right.\\\nonumber&+&\left.\left.9216 \pi ^4 (2 \eta
+3) y^6 \zeta(y) \vartheta (y)^5+4608 \pi ^4 (4 \eta +3) y^6
\vartheta (y)^6\right)+18 \gamma ^2 \left(2 \zeta(y) \vartheta (y)
\left(2 \eta +12 \omega _q-1\right)-12
\zeta(y)^2\right.\right.\\\nonumber&+&\left.\left.\vartheta (y)^2
\left(4 \eta -3 \omega _q \left(6 \omega
_q+1\right)\right)\right)\right).
\end{eqnarray}

It is interesting to mention that $\Omega(y_0)=0$ and $
\Omega^\prime(y_0)=0$. Hence, we have
\begin{equation}
\Omega(y)=
\frac{1}{2}\Omega^{\prime\prime}(y_0)(y-y_0)^2+\mathcal{O}[(y-y_0)^3]\,,
\end{equation}
consequently, the equation of motion becomes
\begin{equation}
\dot
y^2=-\frac{1}{2}\Omega^{\prime\prime}(y_0)(y-y_0)^2+\mathcal{O}[(y-y_0)^3]\,.
\end{equation}

Thus, the developed structure becomes stable if
$\Omega^{\prime\prime}(y_0)> 0$ otherwise unstable. Hence, the
stability condition of the developed structure can be characterized
as
\begin{equation}\label{30}
\eta_0 >\frac{\mathcal{A}_0}{\mathcal{B}_0}\quad \text{if}\quad
\mathcal{B}_0>0, \quad\quad \eta_0
<\frac{\mathcal{A}_0}{\mathcal{B}_0}\quad \text{if}\quad
\mathcal{B}_0<0,
\end{equation}
where
\begin{eqnarray}\nonumber
\mathcal{A}_0&=&3 \gamma  y_0^{3 \omega _q} \left(24 \zeta(y_0)^2
\left(-16 \pi y_0^3 \rho _c+3 \text{C}_2 y_0^2+12
m\right)+\mathcal{D}_1+3 \mathcal{D}_2 \vartheta
(y_0)^2\right)+y_0^{6 \omega _q} \left(6 \zeta(y_0)^2
\left(\left(-16 \pi  y_0^3 \rho _c+3 \text{C}_2 y_0^2+12
m\right){}^2\right.\right.\\\nonumber&+&\left.\left.3072 \pi ^4
y_0^6 \vartheta (y_0)^4\right)+3 y_0^2 \vartheta (y_0)^2 \left(128
\pi ^2 y_0 \vartheta (y_0)^2 \left(4 \pi y_0^3 \left(\rho _c+9 \pi
\vartheta (y_0)^2\right)+3
m\right)+\mathcal{D}_5\right)\right.\\\nonumber&+&\left.\zeta(y_0)
\vartheta (y_0) \left(\mathcal{D}_3+27648 \pi ^4 y_0^6 \vartheta
(y_0)^4\right)\right)+18 \gamma ^2 \mathcal{D}_4,\\\nonumber
\mathcal{B}_0&=&2 \vartheta (y_0) (\zeta(y_0)+\vartheta (y_0))
\left(y_0^{3 \omega _q} \left(16 \pi y_0^3 \left(6 \pi  \vartheta
(y_0)^2-\rho _c\right)+3 \left(\text{C}_2 y_0^2+4 m\right)\right)+6
\gamma \right)\\\nonumber&\times& \left(y_0^{3 \omega _q} \left(3
\left(\text{C}_2 y_0^2+4 m\right)-16 \pi  y_0^3 \left(\rho _c+6 \pi
\vartheta (y_0)^2\right)\right)+6 \gamma \right),
\end{eqnarray}
where
\begin{eqnarray}\nonumber
\mathcal{D}_1&=&4 \zeta(y_0) \vartheta (y_0) \left(16 \pi  y_0^3
\rho _c \left(6 \omega _q-7\right)+3 \left(-6 \omega _q
\left(\text{C}_2 y_0^2+4 m\right)+5 \text{C}_2 y_0^2+4
m\right)\right),\\\nonumber \mathcal{D}_2&=&-16 \pi  y_0^3 \rho _c
\left(\omega _q-1\right) \left(3 \omega _q-2\right)+3 \omega _q
\left(3 \omega _q \left(\text{C}_2 y_0^2+4 m\right)-3 \text{C}_2
y_0^2+4 m\right)+2 \text{C}_2 y_0^2+32 \pi ^2 y_0^3 \left(9 \omega
_q \left(\omega _q+1\right)+2\right) \vartheta (y_0)^2,\\\nonumber
\mathcal{D}_3&=&\left(-208 \pi  y_0^3 \rho _c+27 \text{C}_2 y_0^2+12
m\right) \left(-16 \pi  y_0^3 \rho _c+3 \text{C}_2 y_0^2+12
m\right),\\\nonumber \mathcal{D}_4&=&2 \zeta(y_0) \left(1-12 \omega
_q\right) \vartheta (y_0)+12 \zeta(y_0)^2+3 \omega _q \left(6 \omega
_q+1\right) \vartheta (y_0)^2,\\\nonumber \mathcal{D}_5&=&32 \pi y_0
\rho _c \left(20 \pi  y_0^3 \rho _c-5 \text{C}_2 y_0^2-6 m\right)+3
\text{C}_2 \left(3 \text{C}_2 y_0^2+4 m\right) .
\end{eqnarray}

\begin{figure}\centering
\epsfig{file=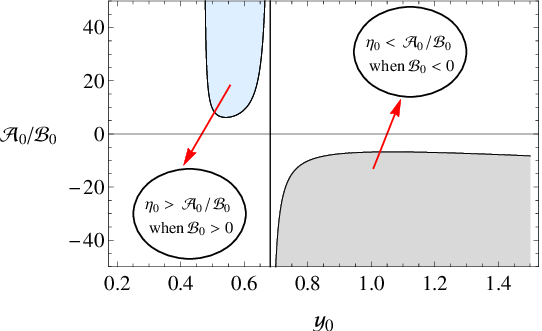,width=.5\linewidth}\epsfig{file=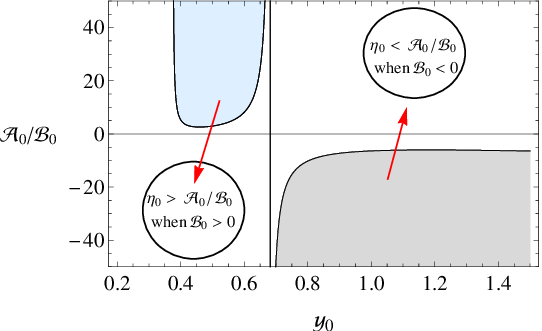,width=.512\linewidth}
\epsfig{file=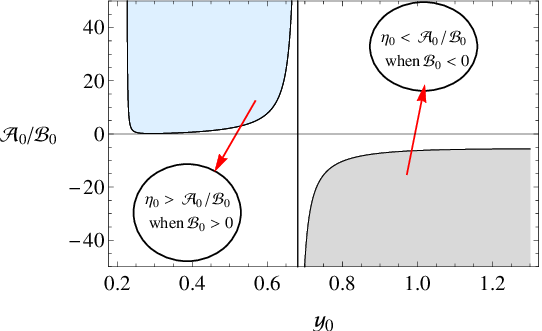,width=.5\linewidth}\epsfig{file=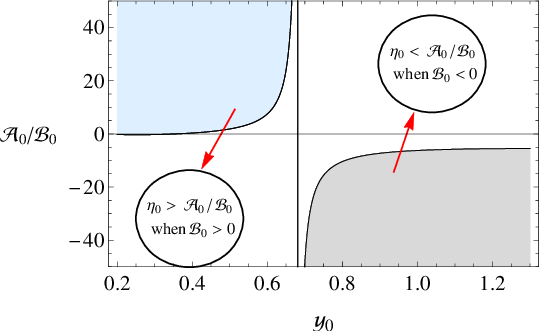,width=.51\linewidth}
\caption{\label{F1} Plots of $\mathcal{A}_0/\mathcal{B}_0$ versus
$y_0$ for different values of cloud parameter $a=0.3$ (first plot),
$a=0.6$ (second plot), $a=0.9$ (third plot), $a=1.2$ (fourth plot)
with $\omega _q=-2/3,\rho _c=0.5,C_2=1,\gamma =0.2,m=0.5$. Here, the
shaded regions depict the stability of gravastars through the
behavior of $\eta_0$ if $\mathcal{B}_0>0$  (light blue region) and
$\mathcal{B}_0<0$ (light gray region).}
\end{figure}
\begin{figure}\centering
\epsfig{file=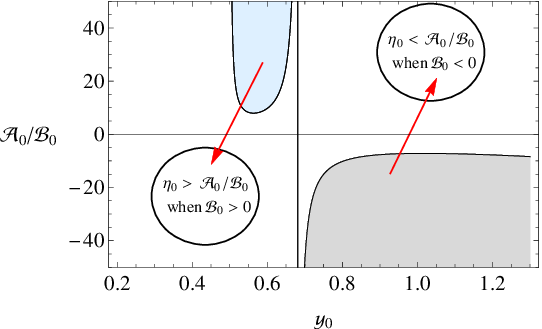,width=.5\linewidth}\epsfig{file=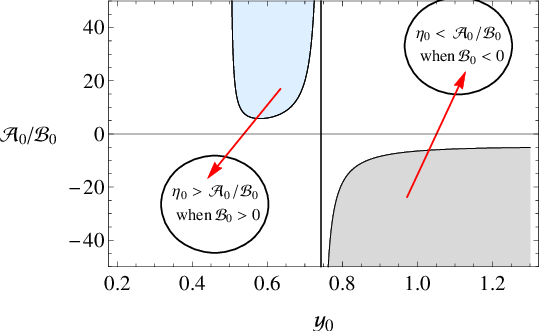,width=.512\linewidth}
\epsfig{file=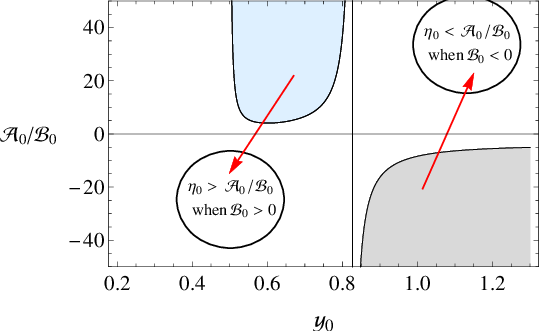,width=.5\linewidth}\epsfig{file=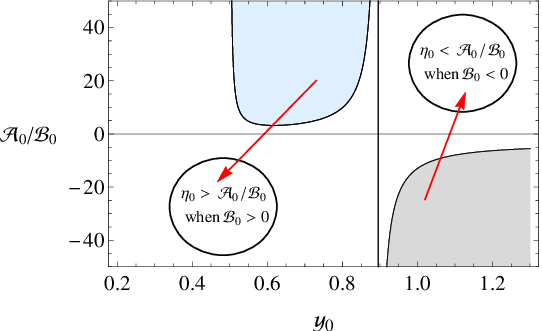,width=.51\linewidth}
\caption{\label{F2} Plots of $\mathcal{A}_0/\mathcal{B}_0$ versus
$y_0$ for different values of quintessence parameter $\gamma=0.2$
(first plot), $\gamma=0.8$ (second plot), $\gamma=1.5$ (third plot),
$\gamma=2$ (fourth plot) with $\omega _q=-2/3,\rho
_c=0.5,C_2=1,a=0.2,m=0.5$.}
\end{figure}

In Figs. (\ref{F1}) and (\ref{F2}), we are interested in exploring
the stable configuration of the gravastars in the framework of
quintessence and cloud of strings. It is very interesting to mention
that the stable regions of the gravastars are greatly affected by
the presence of cloud of strings and quintessence field parameters.
 {Figs. (\ref{F1}) is used to explore the effects of
the cloud of strings parameter on the stability of the gravastars.
It is found that stable regions increase as the cloud of strings
parameter increases. This shows that the developed structure is more
stable due to the effects of the cloud of strings (Figs.
(\ref{F1})). In Figs. (\ref{F2}), we are interested in discussing
the stability of the developed structure by considering different
values of the quintessence field parameter.} It is found that the
stability regions decrease as the quintessence field parameter
increases as shown in (Figs. (\ref{F2})). Hence, both quintessence
and cloud of strings play remarkable role to maintain the stability
of gravatars.

\subsection{Proper Length of the Thin Shell}

We are interested to discuss the proper length of the shell and the
thickness of the shell is represented by $\delta$. The shell
thickness is a very small positive real number such as
$0<\delta\ll1$. The lower and upper boundaries of the shell are $y$
and $y+\delta$. Mathematically, the proper length of the shell can
be evaluated as \cite{qm}
\begin{equation}\label{19}
l=\int_{y}^{y+\delta}\sqrt{e^{\varepsilon(r)}}=\int_{y}^{y+\delta}
\frac{dr}{\sqrt{-\alpha +3^{2/3} e^{-C_{4}} r^{2/3}+1}}.
\end{equation}
In order to solve the above-complicated integration, we assume that
the $\frac{1}{\sqrt{-\alpha +3^{2/3} e^{-C_{4}}
r^{2/3}+1}}=\frac{dW(r)}{dr}$ as
\begin{equation}\label{20}
l=\int_{y}^{y+\delta}\frac{dW(r)}{dr}dr=W(y+\delta)-W(y)\approx
\delta \frac{dW(r)}{dr}|_{r=y}=\delta(-a +3^{2/3} e^{-C_{4}}
y^{2/3}+1)^{-1/2},
\end{equation}
where $\delta\ll1$ is a very small positive real constant,
therefore, its square and higher powers must be ignored. In this
regard, we obtain a relationship between the thickness and the
proper length of the shell. This relation depends on the shell
radius and cloud of strings parameter. The behavior of proper length
versus thickness of the shell for different values of the cloud of
quintessence is shown in the left plot of Fig. (\ref{p}). Proper
length increases as thickness as well as the cloud of strings
parameter increases.

\subsection{Energy content}

The inner region of gravastar is a negative energy zone with a
non-attractive force and matter obeys the EoS $p=-\sigma$.  Applying
the same technique, such as determining the proper length, the
energy distribution in the shell's region can be evaluated as
\cite{qm}
\begin{equation}\label{25}
E=\int_{y}^{y+\delta}4\pi r^2 \rho(r)dr\approx4\delta\pi y^2 C_6
e^{-\frac{3\ 3^{2/3} e^{-C_3} y^{2/3}}{a -1}}.
\end{equation}
The final expression of the shell energy is directly related to the
shell radius, thickness and the cloud of strings parameter. Hence,
the cloud of strings parameter directly effects the energy of the
shell. The graphical analysis of energy content versus thickness of
the shell for different values of quintessence is shown in the
middle plot of Fig. (\ref{p}). It increases as the thickness as well
as the cloud of strings parameter increases.

\subsection{Entropy}

\begin{figure}\centering
\epsfig{file=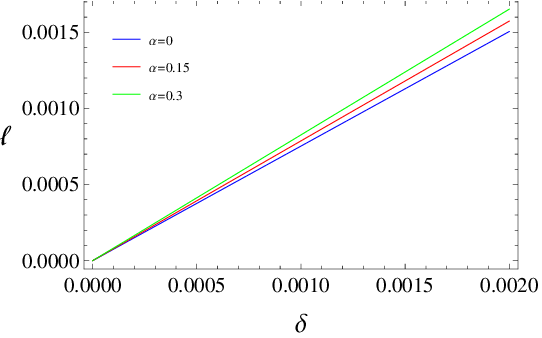,width=.5\linewidth}\epsfig{file=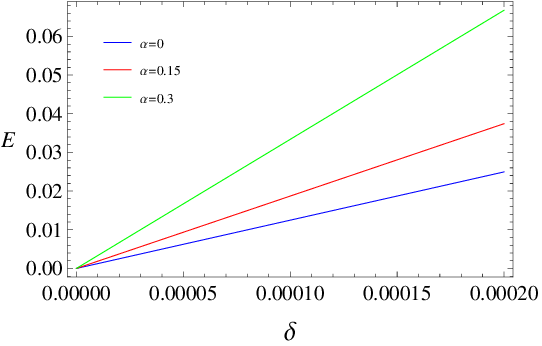,width=.5\linewidth} \epsfig{file=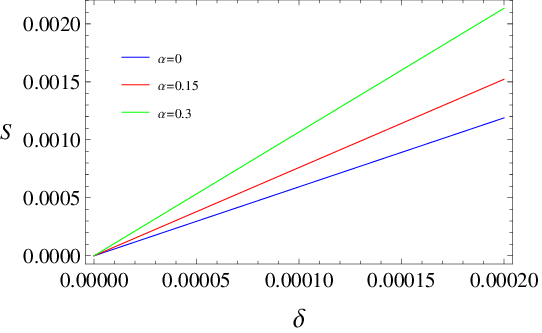,width=.5\linewidth}
\caption{\label{p} Behavior of proper length (left plot), energy
contents (middle plot) and entropy of shell (right plot) for
different values of the cloud of strings parameter $a=0,0.15,0.3$.}
\end{figure}

The degree of disorder or disturbance in a geometric structure is
related to the entropy measure. To understand the randomness of
gravastar geometry, we look at the entropy of thin-shell gravastars.
Mazur and Mottola's idea is used to calculate an equation for the
entropy of a thin-shell gravastar as \cite{qm}
\begin{equation}\label{22}
S=\int_{y}^{y+\delta}4\pi r^2 h(r) \sqrt{e^{\varepsilon(r)}}dr.
\end{equation}
For local temperature, the entropy density is calculated as
\begin{equation}\label{23}
h(r)=\frac{\eta K_B}{\hbar}\sqrt{\frac{p(r)}{2\pi}},
\end{equation}
where $\eta$ is represented as a dimensionless parameter. Here, we
take Planck units $(K_B = 1 = \hbar)$ so that the shell's entropy
becomes \cite{qm}
\begin{equation}\label{24}
S=2 \sqrt{2 \pi } \delta  \eta  y^2 \sqrt{\frac{1}{-a +3^{2/3}
e^{-C_4} y^{2/3}+1}} \sqrt{C_6 e^{-\frac{3\ 3^{2/3} e^{-C_3}
y^{2/3}}{a -1}}}.
\end{equation}
It is noted that shell's entropy is also proportional to $\delta$.
Similarly, we investigate the entropy of the shell along the
thickness of the shell for different values of the cloud of strings
are shown in the right plot of Fig. (\ref{p}). It is noted that the
entropy increases by increasing $\delta$ as well as the cloud of
strings parameter.

\section{Conclusion}\label{sec:6}

The motivation behind the gravastar solution with a modified matter
source in the background of a cloud of strings and quintessence is
to explore alternative models of gravastars and study their
implications in the context of string theory and dark energy.
Gravastars are hypothetical objects that have been proposed as an
alternative to black holes. They are thought to be made up of exotic
matter that can prevent the formation of an event horizon, which is
a defining feature of black holes. Instead, gravastars have a
surface called a ``gravitational vacuum star" that can mimic some of
the properties of a black hole without the singularity at its
center. In this particular scenario, the background is assumed to be
a cloud of strings. String theory is a theoretical framework that
attempts to reconcile quantum mechanics and general relativity by
describing fundamental particles as tiny, vibrating strings. The
cloud of strings provides a unique environment for studying the
behavior of gravastars and their interaction with the underlying
string structure \cite{54q,55q}. Additionally, quintessence is a
form of dark energy that is hypothesized to explain the accelerating
expansion of the universe \cite{42q}-\cite{45q}. By incorporating
quintessence into the gravastar solution, researchers aim to
investigate the potential interplay between exotic matter, string
theory, and dark energy. Overall, the motivation for studying the
gravastar solution with a modified matter source in the background
of a cloud of strings and quintessence is to explore novel
theoretical frameworks, investigate alternative models of black
hole-like objects, and gain a deeper understanding of the
fundamental nature of the universe. For this purpose, we developed
the Einstein field equation in the framework of a modified matter
source. Further, we have calculated the gravastar structure and
their physical properties as mentioned below:

\begin{itemize}

 \item  \textbf{Inner region:}  By applying the EoS to the interior region and analyzing the equations of motion and conservation equation, it has been established that the solution does not have a singularity. Additionally, the energy density and pressure of the system remain constant, which is consistent with the characteristics associated with dark energy.
 \item \textbf{Intermediate thin shell:} We have considered the EoS that follows the intermediate shell condition and also determined the respective metric potential.

 \item \textbf{Outer region:} We have used the exact black hole solution surrounded by a cloud of strings and quintessence as an outer manifold.  {Then, we developed the gravastar structure by considering Darmois-Israel formalism. Then, we determined the components of the stress-energy tensor that play remarkable role in exploring the stability of the gravastars through linearized radial perturbation.} The detailed outcomes related to stability and physical properties of gravastars are given below:

\begin{enumerate}

\item \textbf{Stability analysis:} We are interested in exploring the effects of the cloud of strings and quintessence on the stability of the gravastar structure through linearized perturbation.  {The stable regions of the developed structure are observed in the shaded regions. Here, the straight line represents the position of the event horizon. The stable regions must exit near the expected position of event horizon of the outer black hole spacetime.  It is noted that the cloud of strings parameter enhances the stable regions of gravastars and stability decreases as $a$ decreases, as shown in Fig. (\ref{F1}).} The quintessence field parameter also effects the stability of the gravastars. The stable region decreases as the quintessence field parameter increases (see Fig.
(\ref{F2})).

\item \textbf{Proper length:} We have established a connection between the thickness of the shell and its proper length. This connection is influenced by both the radius of the shell and the parameter representing the cloud of strings. The left plot in Fig. (\ref{p}) illustrates how the proper length changes with varying thickness of the shell for different values of the cloud of quintessence. As both the thickness and the cloud of strings parameter increase, the proper length also increases.

\item \textbf{Energy:} The equation for the energy of the shell is directly dependent on the shell radius, thickness, and the parameter representing the cloud of strings. As a result, the energy of the shell is directly influenced by the cloud of strings parameter. The middle plot in Fig. (\ref{p}) illustrates how the energy content changes with varying thickness of the shell for different values of the cloud of strings. The energy content increases as both the thickness and the cloud of strings parameter increase.

\item \textbf{Entropy:} The analysis reveals that the entropy of the shell's region is directly proportional to the thickness of the shell. Likewise, we examine the entropy of the shell as we vary its thickness for different values of the cloud of strings. This investigation is depicted in the right plot of Fig. (\ref{p}). It is worth noting that the entropy increases as both the thickness and the cloud of strings parameter increase.
\end{enumerate}
\end{itemize}

 {In the presence of a cloud of strings and
quintessence fields, gravastar solutions are determined to possess
desirable properties. An improvement in the cloud of strings
parameter is favorably connected with gravastar stability, but an
increase in the quintessence field parameter is adversely
correlated.} We have developed a singularity-free solution that is
physically acceptable. It is interesting to mention that the
developed exact novel solutions are reduced to the exact Mazur and
Mottola model in the absence of the cloud of strings and
quintessence fields.  {In future, this work can be
extended for the choice of perfect fluid dark matter. It would be
very interesting to explore the physical configuration of gravastars
in the background of both perfect fluid dark matter and cloud of strings by
using conformal approach.}

\section*{Acknowledgement}
F. Javed acknowledges the financial support provided through Grant
No. YS304023917, which has contributed to his Postdoctoral
Fellowship at Zhejiang Normal University.

\end{document}